\documentclass{cimento}

\usepackage{graphicx} 
\usepackage{verbatim}
\usepackage{amsfonts}
\usepackage{color}
\usepackage{epsfig}
\usepackage{amsmath}
\usepackage{braket}
\numberwithin{equation}{section}
\usepackage{nicefrac}

\title{Quantum plasmonic waveguides: Au nanowires}
\author{C.E.A.~Cordaro\from{ins:x}\ETC \thanks{concetto.cordaro@studium.unict.it},
G.~Piccitto\from{ins:y} \atque
F.~Priolo\from{ins:x}\from{ins:y}
}
\instlist{\inst{ins:x} Scuola Superiore di Catania, Universit\`a di Catania, via Valdisavoia 9, 95123 Catania, Italy
  \inst{ins:y} Dipartimento di Fisica e Astronomia, Universit\`a di Catania, via S. Sofia 64, 95123 Catania, Italy}

\begin{document}

\maketitle

\begin{abstract}
Combining miniaturization and good operating speed is a compelling yet crucial task for our society. \textit{Plasmonic waveguides} enable the possibility of carrying information at optical operating speed while maintaining the dimension of the device in the nanometer range. Here we present a theoretical study of plasmonic waveguides extending our investigation to structures so small that Quantum Size Effects (QSE) become non negligible, namely \textit{quantum plasmonic waveguides}. Specifically, we demonstrate and evaluate a blue-shift in Surface Plasmon (SP) resonance energy for an ultra-thin gold nanowire. 
\end{abstract}

\section{Introduction}
It is well known that the \textit{diffraction limit} imposes a lower bound to the diameter of a common dielectric optical waveguide. Plasmonics indeed gives us the opportunity to confine light at a nanometer scale in metal waveguides via \textit{Surface Plasmon Polaritons} (SPPs). The purpose of the second section is the theoretical analysis of SPPs at the interface between a cylindrical conductor (nanowire) and a dielectric medium. In the third section we consider the case of a nanowire with a radius small enough to show QSE.

\section{SPPs on conducting nanowires}
In the following section Maxwell's equations in cylindrical coordinates are solved and a transcendental equation is found by applying the appropriate boundary conditions. The solutions are the dispersion relations for the various modes of an SPP propagating along a conducting nanowire. First, let us define the parameters concerning the system: we take the axis of the nanowire (of radius \textit{R}) to be the \textit{z} direction and $\epsilon_1$, $\epsilon_2$ to be the permittivity of the metal and of the surrounding dielectric medium respectively. Moreover, we assume a time and \textit{z} dependence of the fields of the form $e^{i({\omega}t-{\beta}z)}$ which implies $\nicefrac{\partial}{\partial t} \to i\omega$ and $\nicefrac{\partial}{\partial z} \to -i\beta$. We then separate Maxwell's curl equations into components solving for $E_x, E_y, H_x$ and $H_y$ in terms of $E_z, H_z$. For the latter we use Helmholtz equations instead.
Switching to cylindrical coordinates $(r,\theta,z)$ we obtain \footnote{Detailed steps can be easily found in several textbooks as \cite{ramo2008fields}}
{\small \begin{equation}
	\begin{split}
		E_x &= \frac{i}{\gamma^2}\left[\beta\frac{{\partial}E_z}{{\partial}x} + \omega\mu_0\frac{{\partial}H_z}{{\partial}y}\right]  \\
		E_y &= \frac{i}{\gamma^2}\left[\beta\frac{{\partial}E_z}{{\partial}y} - \omega\mu_0\frac{{\partial}H_z}{{\partial}x}\right]\\
		H_x &= \frac{i}{\gamma^2}\left[\beta\frac{{\partial}H_z}{{\partial}x} - \omega\epsilon_0\epsilon\frac{{\partial}E_z}{{\partial}y}\right]\\
		H_y &= \frac{i}{\gamma^2}\left[\beta\frac{{\partial}H_z}{{\partial}y} + \omega\epsilon_0\epsilon\frac{{\partial}E_z}{{\partial}x}\right]\\
		\nabla^2_tE_z &= \gamma^2E_z\\
		\nabla^2_tH_z &= \gamma^2H_z
	\end{split}
	\quad\rightarrow\quad
	\begin{split}
		E_r &= \frac{i}{\gamma^2}\left[\beta\frac{{\partial}E_z}{{\partial}r} + \frac{\omega\mu_0}{r}\frac{{\partial}H_z}{{\partial}\theta}\right]\\
		E_{\theta} &= \frac{i}{\gamma^2}\left[\frac{\beta}{r}\frac{{\partial}E_z}{{\partial}\theta} - \omega\mu_0\frac{{\partial}H_z}{{\partial}r}\right]\\
		H_r &= \frac{i}{\gamma^2}\left[\beta\frac{{\partial}H_z}{{\partial}r} - \frac{\omega\epsilon_0\epsilon}{r}\frac{{\partial}E_z}{{\partial}\theta}\right]\\
		H_{\theta} &=  \frac{i}{\gamma^2}\left[\frac{\beta}{r}\frac{{\partial}H_z}{{\partial}\theta} + \omega\epsilon_0\epsilon\frac{{\partial}E_z}{{\partial}r}\right]\\
		 \gamma^2E_z &= \frac{1}{r}\frac{\partial}{{\partial}r}\left(r\frac{{\partial}E_z}{{\partial}r} \right) + \frac{1}{r^2}\frac{{\partial^2}E_z}{{\partial}\theta^2}\\
		\gamma^2H_z &= \frac{1}{r}\frac{\partial}{{\partial}r}\left(r\frac{{\partial}H_z}{{\partial}r} \right) + \frac{1}{r^2}\frac{{\partial^2}H_z}{{\partial}\theta^2}  
	\end{split}
\label{maxset}
\end{equation}}
where $\gamma^2=\beta^2-k^2=\beta^2-\omega^2\mu_0\epsilon\epsilon_0$ and $\mu=1$. The last two differential equations for $E_z$ and $H_z$ are separable in $r$ and $\theta$. Furthermore, the part describing the $r$ dependence is in the form of the \textit{modified Bessel differential equation} leading to the general solution for $E_z$ and $H_z$ 
\begin{equation}
\left\lbrace A I_{\nu}({\gamma}r) + B K_\nu({\gamma}r)\right\rbrace \left\lbrace C\cos{\nu\theta} + D\sin{\nu\theta}\right\rbrace{e^{i({\omega}t - {\beta}z)}} 
\label{eq:Bessel7}
\end{equation}
where $A, B, C$ and $D$ are constants and $I_\nu$ and $K_\nu$ are the $\nu$-th order modified Bessel functions of the first and second kind respectively. Due to their behavior at $r=0$ and $r\to\infty$, $K_\nu$ inside the core ($r<R$) and $I_\nu$ outside ($r>R$) must be excluded \cite{stratton2007}. Hence, for the order $\nu$:
{\small 
	\begin{equation}
	\begin{split}
\begin{cases}
	\mbox{~}\mbox{if}& r<R\\
	\mbox{~}E_{z1}&=CI_{\nu}({\gamma}r)\cos{(\nu\theta)}{e^{i({\omega}t - {\beta}z)}}  \\
	\mbox{~}H_{z1}&=C'I_\nu({\gamma}r)\cos{(\nu\theta)}{e^{i({\omega}t - {\beta}z)}}
\end{cases}
	\end{split}
	\qquad
	\begin{split}
	\begin{cases}
	\mbox{~}\mbox{if}& r>R\\
	\mbox{~}E_{z2}&=DK_{\nu}({\gamma}r)\cos{(\nu\theta)}{e^{i({\omega}t - {\beta}z)}}\\
	\mbox{~}H_{z2}&=D'K_\nu({\gamma}r)\cos{(\nu\theta)}{e^{i({\omega}t - {\beta}z)}} \\
	\end{cases}
	\end{split}
	\label{fieldsSPP}
	\end{equation}}
where $C, C', D$ and $D'$ are constants to be determined. The other components of the fields can be easily computed inserting eqs.(\ref{fieldsSPP}) back in eqs.(\ref{maxset}). 

\begin{figure}
	\centering
	\includegraphics[width=.47\textwidth]{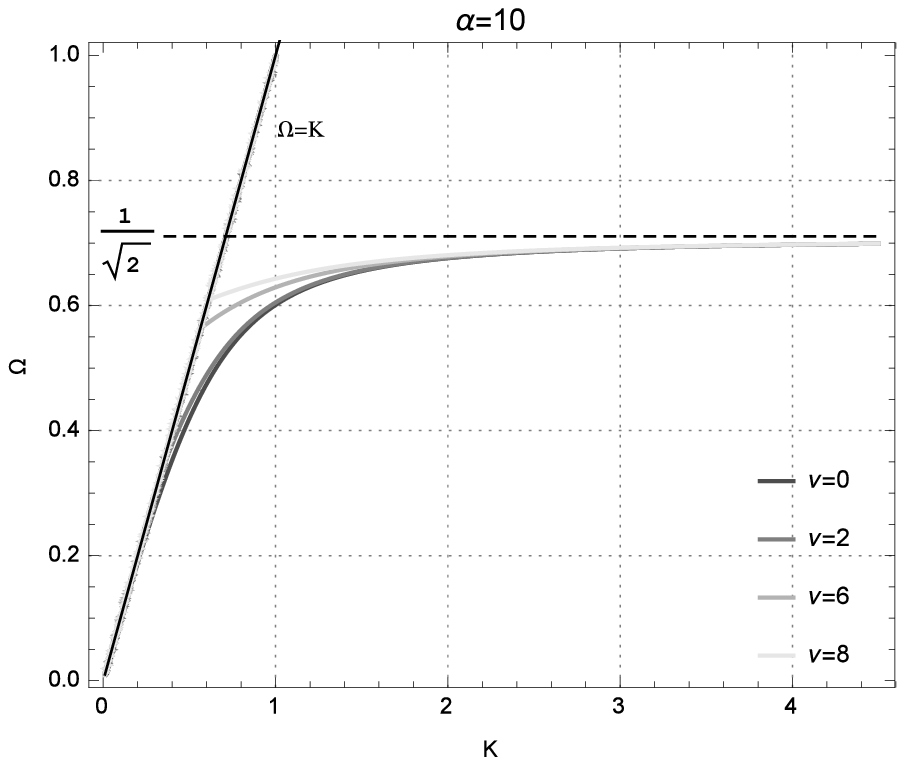}\hspace{0.5cm}
	\includegraphics[width=.47\textwidth]{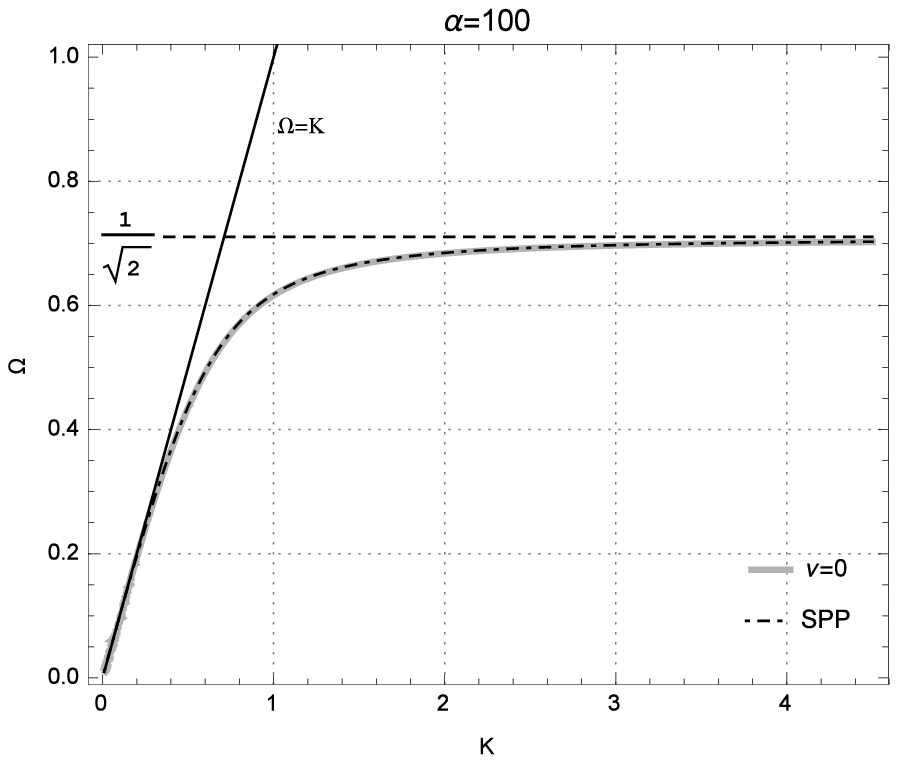}
	\caption{\textbf{Left}: dispersion relation for a SPP on a conducting wire for $\alpha=10$ concerning modes $\nu=0, 2, 6, 8$. \textbf{Right}: comparison between TM mode for a SPP on a conducting wire and a SPP on a planar interface (dot-dashed line)}
	\label{fig:disp2}
\end{figure}

Imposing the continuity of the tangential components of both fields at $r=R$ results in a homogeneous system of four linear equations with $C, C', D$ and $D'$ as unknowns \cite{stratton2007}. This system admits no solution for $C=D=0$ or $C'=D'=0$ with the only exception of the case $\nu=0$ for which the first solution is a \textit{TE mode}\footnote{TE (TM) waves: there is no electric (magnetic) field in the direction of propagation.} (which is not a propagating mode for the system under investigation \cite{Takahara:04}) and the second solution is a \textit{TM mode}. For $\nu>0$ all modes are hybrid ($E_z\neq0$ and $H_z\neq0$). In order to find nontrivial solutions, the determinant of the system must be set to zero leading to the following transcendental equation whose solutions are the dispersion relations $\omega_\nu(\beta)$ for mode $\nu$
{\small \begin{equation}
\left[\frac{1}{\xi_1}\frac{I'_{\nu}(\xi_1)}{I_{\nu}(\xi_1)} - \frac{1}{\xi_2}\frac{K'_{\nu}(\xi_2)}{K_{\nu}(\xi_2)}\right]  \left[\frac{\epsilon_1}{\xi_1}\frac{I'_{\nu}(\xi_1)}{I_{\nu}(\xi_1)} - \frac{\epsilon_2}{\xi_2}\frac{K'_{\nu}(\xi_2)}{K_{\nu}(\xi_2)}\right] = \nu^2\left(\frac{c\beta}{\omega}\right)^2\left(\frac{1}{\xi^2_1}-\frac{1}{\xi^2_2}\right)^2 \\
\label{eq:caratteristica}
\end{equation}}
where $\gamma_j=\sqrt{\beta^2-\epsilon_j\left(\nicefrac{\omega}{c}\right)^2}$, $\xi_j=\gamma_jR$ and $j=1, 2$.
\begin{figure}[t]
	\centering
	\includegraphics[width=.46\textwidth]{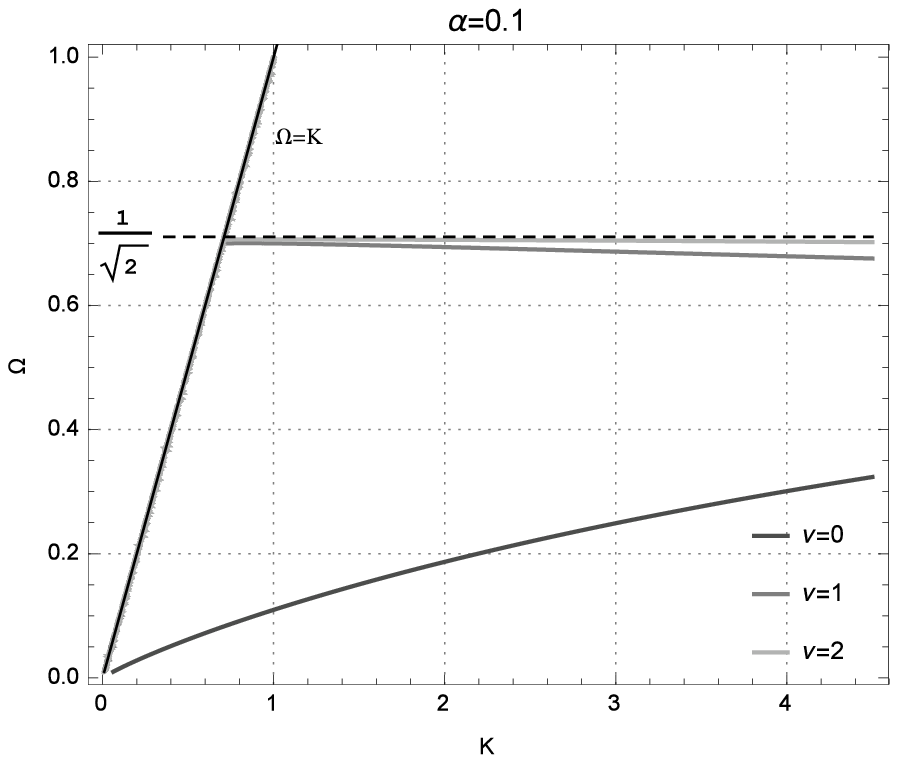}\hspace{0.5cm}
	\includegraphics[width=.46\textwidth]{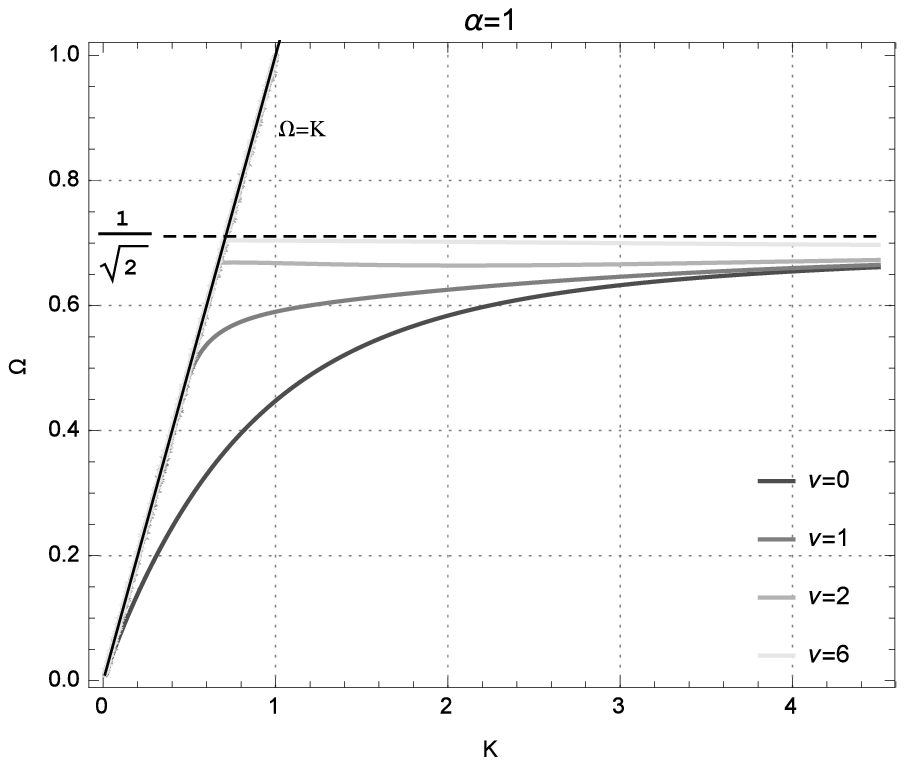}
	\caption{Dispersion relations for a SPP on a conducting wire for of $\alpha=0.1, 1$ concerning modes $\nu=0, 1, 2, 6$.}
	\label{fig:disp1}
\end{figure}
Before embarking on solving eq.(\ref{eq:caratteristica}), it is very useful to perform the change of variable $\Omega = \nicefrac{\omega}{\omega_p}$, $K = \nicefrac{\beta}{k_p}=\nicefrac{c\beta}{\omega_p}$ and to define the \textit{cylindricality constant} $\alpha=\nicefrac{R\omega_p}{c}$ \cite{PhysRevB.10.3038}. The advantage is that $\Omega$, $K$ and $\alpha$ are dimensionless and the new dispersion relations $\Omega_\nu=\Omega_\nu(K)$ do not depend on the the particular value of $\omega_p$. We also assume $\epsilon_1=1-\nicefrac{\omega^2_p}{\omega^2}$ (where $\omega_p$ is the plasma frequency for the metal under investigation) and $\epsilon_2=1$. Equation (\ref{eq:caratteristica}) becomes
{\footnotesize 
\begin{multline*}
\bigg[ \frac{I'_{\nu}(\alpha\sqrt{K^2-\Omega^2+1})}{\alpha\sqrt{K^2-\Omega^2+1}\cdot[I_{\nu}(\alpha\sqrt{K^2-\Omega^2+1})]}- \frac{K'_{\nu}(\alpha\sqrt{K^2-\Omega^2})}{\alpha\sqrt{K^2-\Omega^2}\cdot[K_{\nu}(\alpha\sqrt{K^2-\Omega^2})]}
\bigg]\cdot 
\bigg[
\frac{1-\nicefrac{1}{\Omega^2}}{\alpha\sqrt{K^2-\Omega^2+1}}\cdot\\
\frac{I'_{\nu}(\alpha\sqrt{K^2-\Omega^2+1})}{I_{\nu}(\alpha\sqrt{K^2-\Omega^2+1})}-\frac{K'_{\nu}(\alpha\sqrt{K^2-\Omega^2})}{\alpha\sqrt{K^2-\Omega^2}\cdot[K_{\nu}(\alpha\sqrt{K^2-\Omega^2})]}
\bigg]
-\bigg[\frac{\nu K}{\alpha\omega}\bigg]^2\bigg[\frac{1}{K^2-\Omega^2+1}-\frac{1}{K^2-\Omega^2}\bigg]^2=0
\label{eq:caratteristica2}
\end{multline*} }
Results obtained\footnote{From a merely computational point of view, we found more convenient to treat the left-hand side of the equation as a function $F$ of the variables ($K$,$\Omega$) hence obtaining the dispersion relations by imposing $F(K,\Omega)=0$. We also used $I'_{\nu}=\frac{1}{2}\left(I_{\nu-1}+I_{\nu+1}\right)$ and $K'_{\nu}=-\frac{1}{2}\left(K_{\nu-1}+K_{\nu+1}\right)$.} for modes $\nu=0, 1, 2, 6, 8$ as the parameter $\alpha$ changes from $0.1$ to $100$ are shown in figs.\ref{fig:disp1} and \ref{fig:disp2}. The dashed line $\Omega=\nicefrac{1}{\sqrt{2}}$ denotes the frequency of the \textit{non-retarded surface plasmon mode} $\omega_{sp}=\nicefrac{\omega_{p}}{\sqrt{2}}$ while the straight solid line $\Omega=K$ is the dielectric light line. The behavior at $K\to\infty$ is common to each mode as $\alpha$ changes: the frequency approaches $\omega_{sp}$ and the group velocity $\nicefrac{d\Omega}{dK}\to0$ thus implying an electrostatic character.
It is worth remarking that the electrostatic solution's frequency $\omega_{sp}$ for a SPP on a conducting wire is the same as that on a planar metal-dielectric interface. \\
Concerning the dependence on the parameter $\alpha$, it is easy to observe that for high cylindricality ($\alpha\ll1$) the frequency of the modes $\nu>0$ is very close to that of the surface plasmon. As the cylindricality decreases ($\alpha$ increases), all these modes start collapsing on the first $\nu=0$ mode. The latter is a very peculiar mode: looking at eqs.(\ref{fieldsSPP}) we notice that both the fields depend on $\theta$ only trough $\cos{(\nu\theta)}$, hence for $\nu=0$ the SPP sees no curvature and it behaves as if it were on a planar interface. This is the reason why all the other modes approach to $\nu=0$ as the radius increases. Moreover, from fig.\ref{fig:disp2} \textbf{right} it is clear that the dispersion relation of TM mode for a SPP on a conducting wire and that of a SPP on a planar interface coincide if the radius is sufficiently large. 

\section{Quantum Size Effects}
\begin{figure}
	\centering
	\includegraphics[width=.516\textwidth]{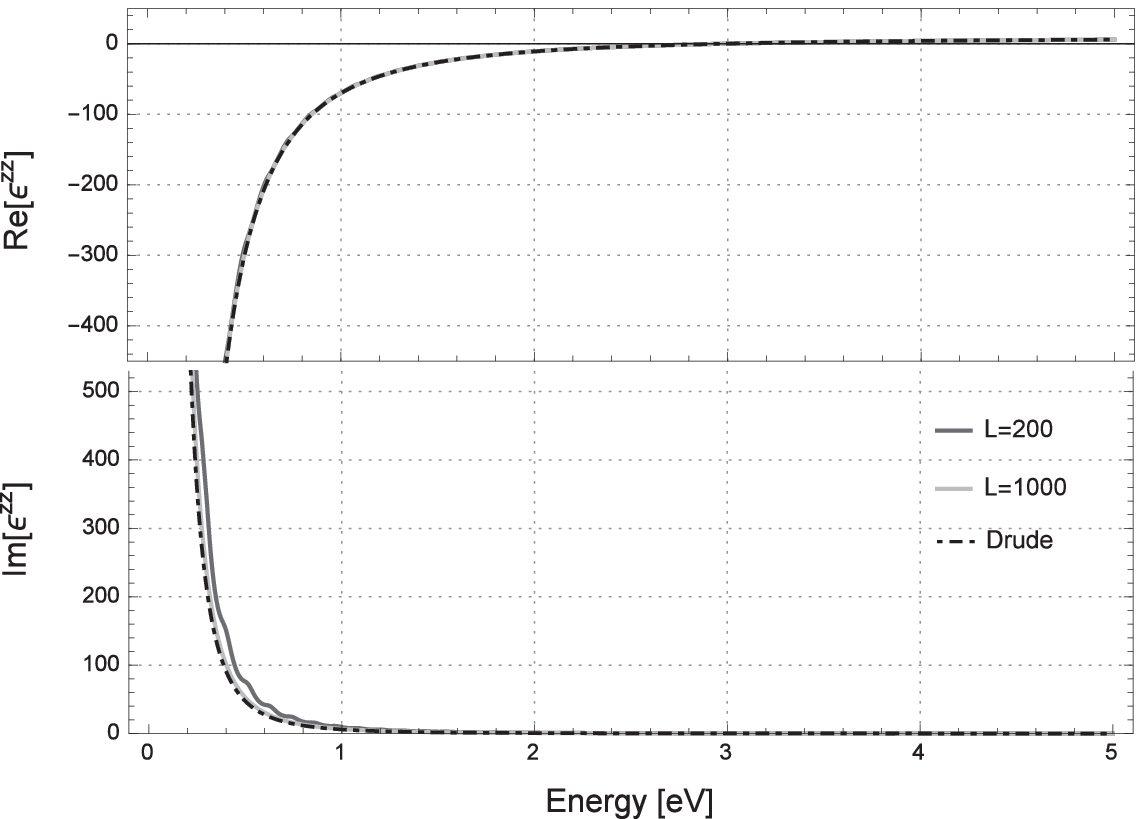}\hfil
	\includegraphics[width=.484\textwidth]{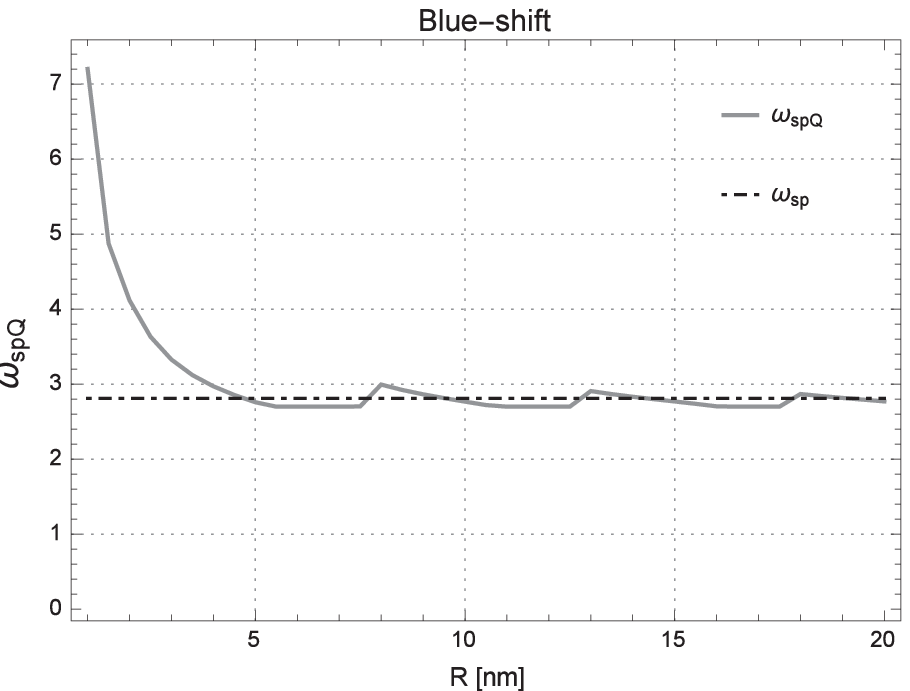}
	\caption{\textbf{Left}: real (\textbf{up}) and imaginary (\textbf{down}) part of $\epsilon^{zz}$ as $L$ changes. \textbf{Right}: $\omega_{spQ}$ as a function of $R$}
	\label{fig:rezz}
\end{figure}

So far, the problem of a SPP propagating along a metallic nanowire has been treated classically. However, as Fr{\"o}lich noticed back in 1937 \cite{frohlich1937spezifische}, when the size of the nanostructure sustaining SPPs becomes small enough, the continuous electronic conduction band breaks up into discrete states and \textit{Quantum Size Effects} (QSE) arise. As a consequence, the simple Drude model used in the previous section is no longer valid and must be corrected in a quantum manner.
One specific way to handle the problem\cite{genzel1975} consists in treating the system as a non-interacting free electron gas constrained by the physical boundaries of the nanostructure. These electrons can undergo transitions from occupied states $\Ket{\psi_i}$ lying below the Fermi surface to unoccupied states $\Ket{\psi_f}$ above it, with a characteristic frequency $\omega_{if}$. Accordingly, the dielectric response of the electronic plasma is modeled adding quantum-mechanically derived Lorentzian terms to the Drude model, weighting them with the \textit{oscillator strengths} $S_{if}$ as follows
{\small \begin{equation}
\epsilon^{m,n}(\omega)={\epsilon_{\infty}}+\omega^2_p
\sum_{\substack{i}}\sum_{\substack{f}}\frac{S_{if}}{\omega^2_{if} - \omega^2 - i\Gamma'\omega}
\label{eq:drude-lorentzQuantum}
\end{equation}}
Thanks to the symmetry of our system and taking again the axis of the nanowire (of radius \textit{R}) to be the \textit{z} direction, the permittivity is a diagonal tensor $\bar{\bar\epsilon}= diag\left\lbrace \epsilon^{xx},\epsilon^{xx},\epsilon^{zz}\right\rbrace$. Each component of $\bar{\bar\epsilon}$ is obtained evaluating the sums in eq.(\ref{eq:drude-lorentzQuantum}). We now proceed explaining the terms of the latter while we compute $\epsilon^{xx}$ for a gold nanowire of length \textit{L}. In this case, given the Thomas - Reiche - Kuhn sum rule\footnote{see Appendix $A_{XIII}$ of \cite{claudequantum}}, we have
{\small \begin{equation}
S_{if}=\frac{2M\omega_{if}}{N\hbar}\left|\Bra{\psi_i}x\Ket{\psi_f}\right|^2
\label{eq:Sif}
\end{equation}}
where $\Ket{\psi_i}$ and $\Ket{\psi_f}$ are the eigenfunctions for each of the $N$ particles in a infinite cylindrical quantum well and $\omega_{if}$ is Bohr frequency of the transition 
{\small \begin{equation}
\omega_{if}=\frac{E_f - E_i}{\hbar}
\label{eq:omegaif}
\end{equation}}
where $E_f$ and $E_i$ are the energies of the final and initial state respectively. The damping coefficient $\Gamma'$ appearing in~(\ref{eq:drude-lorentzQuantum}) modifies the bulk counterpart taking into account the size of the nanostructure  
{\small \begin{equation}
\Gamma'= \Gamma + \frac{v_f}{R_{eff}}
\label{eq:gammaif}
\end{equation}}
\begin{figure}
	\centering
	\includegraphics[width=.49\textwidth]{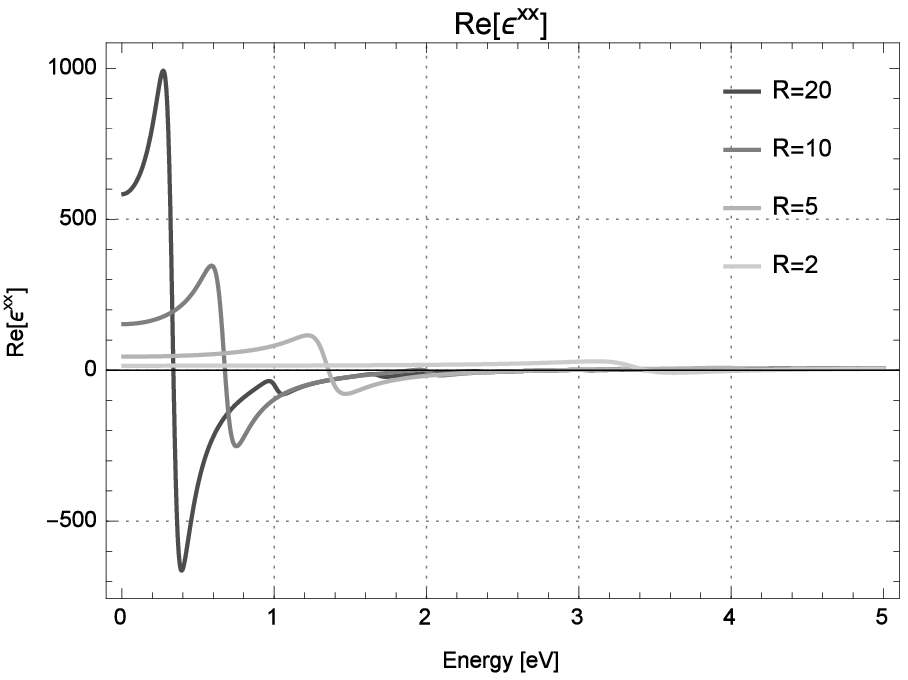}\hfil
	\includegraphics[width=.49\textwidth]{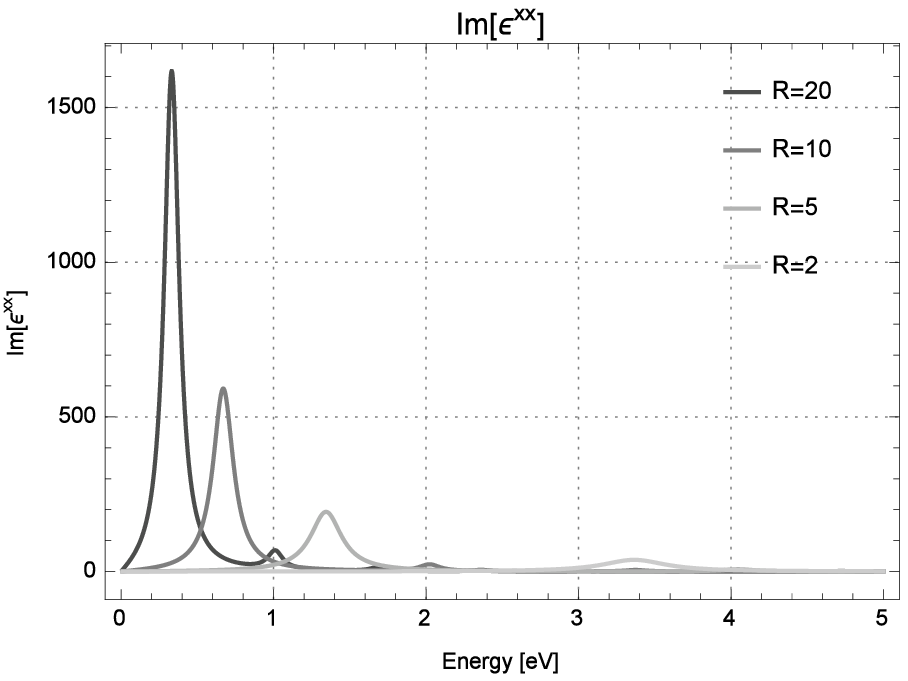}
	\caption{Real (\textbf{left}) and imaginary (\textbf{right}) part of $\epsilon^{xx}$ as $R$ changes}
	\label{fig:rexx}
\end{figure}
where $v_f$ is the Fermi velocity and $R_{eff}$ is an effective radius parameter; for a cylinder $\frac{v_f}{R_{eff}}=\frac{9\pi\left({3\pi}\right)^{1/3}}{64}\frac{v_f}{R}$\cite{Kraus1983}.
The last two terms appearing in~(\ref{eq:drude-lorentzQuantum}) are $\epsilon_{\infty}$ and $\omega_p$: the first is a corrective coefficient that accounts for the behavior of noble metals for $\omega\gg\omega_p$\cite{maier2010plasmonics} and the latter is the gold plasma frequency.\\
The evaluation both of the matrix element in eq.(\ref{eq:Sif}) and of eq.(\ref{eq:omegaif}) using infinite cylindrical quantum well eigenfunctions and energy eigenvalues can be found elsewhere\cite{Kraus1983}. The same holds also for $\epsilon^{zz}$. The actual summations in eq.(\ref{eq:drude-lorentzQuantum}) have been carried numerically. The results obtained\footnote{See \cite{PhysRevB.71.085416} and \cite{scholl2012quantum} for the coefficients and parameters used for gold} for the real and imaginary part of both $\epsilon^{xx}$ and $\epsilon^{zz}$ as \textit{L} and \textit{R} change are shown in figs.\ref{fig:rexx} and \ref{fig:rezz}. It is quite encouraging that if the field is polarized along \textit{z} direction we recover the classical Drude model (see fig.\ref{fig:rezz} \textbf{left}) even though we started from a quantum defined dielectric function. The reason is that along \textit{z} the quantum confinement is not so high hence the spacing between energy levels is very small and the conduction band can be considered continuous again. Conversely, QSE are evident with regard to $\epsilon^{xx}$ (fig.\ref{fig:rexx}).
Changing the dielectric functions of the system certainly affects the SPPs dispersion relation. Specifically, a blue-shift in Surface Plasmon resonance energy occurs and the final result of our work is its quantitative evaluation. The condition that has to be met to have surface plasmon resonance in our system is\cite{raether1988surface}
{\small \begin{equation}
Re\left[ \epsilon^{xx}(\omega_{spQ})\right] = -\epsilon_2 
\label{eq:condizioneSP}
\end{equation}}
We solved eq(\ref{eq:condizioneSP}) numerically for $\omega_{spQ}$ as \textit{R} changed from 1nm to 20nm with $\epsilon_2=1$, as displayed in fig.\ref{fig:rezz} \textbf{right}. The plot clearly shows a consistent blue-shift as the radius of the wire decreases below 5nm. On the other hand, when the radius increases to ``normal'' values, $\omega_{spQ} \to \omega_{sp}=\nicefrac{\omega_p}{\sqrt{1+\epsilon_{\infty}}}$ recovering again the classical situation\footnote{The little peaks of $\omega_{spQ}$ for certain values of \textit{R} are not a computational artifact but are caused instead by smaller secondary oscillations in the dielectric function (both real and imaginary) such that barely visible around $\sim1 eV$ for $R=20 nm$ in fig.\ref{fig:rexx} }.

\section{Discussion}
In the present work we analyzed the plasmonic properties of metal waveguides both from a classical point of view and from a quantum mechanical one. 
We derived and presented the dispersion relations for an SPP on a conducting wire showing in particular the similarity of the $\nu=0$ mode to the SPP on a planar interface and, for large radii, of all the other modes to the $\nu=0$ mode. 
The main result of our work is the demonstration and the quantification of the {\textit{blue-shift in SP resonance}} for a 1D system such an ultra-thin gold nanowire. We used an analytical quantum model accounting for QSE and showed how this would return the classical Drude model if quantum confinement is not high ($\epsilon^{zz}\approx\epsilon_{Drude}$) and the functions of fig.\ref{fig:rexx} if it is. Substituting the dielectric functions eventually resulted in the aforementioned  blue-shift which is also comparable to that experimentally demonstrated by Scholl et al.\cite{scholl2012quantum} for the 0D case of silver nanoparticles.
\acknowledgments
{\small We would like to thank Professor G.G.N. Angilella and Professor L.C. Andreani for their valuable help and constructive comments on our work.
}
\bibliographystyle{varenna}
{\tiny \bibliography{biblio}}

\begin{thebibliography}{10}
\expandafter\ifx\csname url\endcsname\relax\def\url#1{\texttt{#1}}\fi
\expandafter\ifx\csname urlprefix\endcsname\relax\def\urlprefix{URL }\fi

\bibitem{ramo2008fields}
\NAME{Ramo S., Whinnery J.~R. \atque {Van~Duzer} T.}, \TITLE{Fields and Waves
  in Communication Electronics} (Wiley) 1994.

\bibitem{stratton2007}
\NAME{Stratton J.}, \TITLE{Electromagnetic Theory} (Wiley) 2007.

\bibitem{Takahara:04}
\NAME{Takahara J. \atque Kobayashi T.}, \IN{Opt. Photon. News}{15}{2004}{54}.

\bibitem{PhysRevB.10.3038}
\NAME{Pfeiffer C., Economou E. \atque Ngai K.}, \IN{Phys. Rev.
  B}{10}{1974}{3038}.

\bibitem{frohlich1937spezifische}
\NAME{Fr{\"o}hlich H.}, \IN{Physica}{4}{1937}{406}.

\bibitem{genzel1975}
\NAME{Genzel L., Martin T. \atque Kreibig U.}, \IN{Zeitschrift fur Physik B
  Condensed Matter}{21}{1975}{339}.

\bibitem{claudequantum}
\NAME{Cohen-Tannoudji C., Diu B. \atque Laloe F.}, \TITLE{Quantum Mechanics
  Volume 2} (Wiley) 2006.

\bibitem{Kraus1983}
\NAME{Kraus W.~A. \atque Schatz G.~C.}, \IN{The Journal of Chemical
  Physics}{79}{1983}{6130}.

\bibitem{maier2010plasmonics}
\NAME{Maier S.~A.}, \TITLE{Plasmonics: Fundamentals and Applications}
  (Springer) 2010.

\bibitem{PhysRevB.71.085416}
\NAME{Vial A., Grimault A.~S., Macias D., Barchiesi D. \atque de~la Chapelle
  M.~L.}, \IN{Phys. Rev. B}{71}{2005}{085416}.

\bibitem{scholl2012quantum}
\NAME{Scholl J.~A., Koh A.~L. \atque Dionne J.~A.},
  \IN{Nature}{483}{2012}{421}.

\bibitem{raether1988surface}
\NAME{Raether H.}, \TITLE{Surface plasmons on smooth and rough surfaces and on
  gratings}, Springer tracts in modern physics (Springer) 1988.

\end{thebibliography}

\end{document}